\documentclass[graybox]{svmult}
\usepackage{type1cm}
\usepackage{graphicx}
\usepackage{multicol}
\usepackage[bottom]{footmisc}
\usepackage{siunitx} %
\usepackage{derivative} %
\usepackage{mathtools}
\usepackage{amssymb}
\usepackage{lmodern}
\usepackage{microtype}
\usepackage{booktabs}
\usepackage{tabularx}
\usepackage{bm}

\usepackage[style=nature,giveninits=true,sorting=none,useprefix=true]{biblatex}
\usepackage[hidelinks]{hyperref}
\usepackage{cleveref}
\usepackage{fancyhdr}

\renewcommand{\chaptermark}[1]{\markright{#1}}
\renewcommand{\sectionmark}[1]{\markright{#1}}

\renewcommand{\headrulewidth}{0pt}

\newcommand{\rv}{{\mathbf r}}
\newcommand{\Rv}{{\mathbf R}}
\newcommand{\xv}{{\mathbf x}}

\newcommand{\rhob}{\rho_{\mathrm b}}
\newcommand{\rhov}{\bm{\rho}}
\newcommand{\muv}{\bm{\mu}}
\newcommand{\Dcal}{\mathcal{D}}
\newcommand{\Ncal}{\mathcal{N}}
\newcommand{\mlrm}{\mathrm{ML}}

\newcommand{\kb}{k_B}
\newcommand{\rmex}{{\mathrm{ex}}}
\newcommand{\rmext}{{\mathrm{ext}}}

\newcommand{\rmid}{{\mathrm{id}}}
\newcommand{\avg}[1]{\Big\langle {\protect #1} \Big\rangle}
\newcommand{\reals}{\mathbb{R}}
\newcommand{\FE}{\mathcal{F}}
\newcommand{\Fex}{\mathcal{F}_\text{ex}}
\newcommand{\fex}{f_\text{ex}}
\newcommand{\vext}{V^\mathrm{ext}}
\newcommand{\req}{\rho_\mathrm{eq}}
\newcommand{\vv}{{\mathbf v}}
\newcommand{\Jv}{{\mathbf J}}
\newcommand{\Jn}{{\mathbf J}_\text{neq}}
\newcommand{\Pid}{P_\text{id}}
\newcommand{\Pex}{P_\text{ex}}
\newcommand{\dd}{\text{d}}

\DeclareMathOperator{\MSE}{MSE}
\newcommand{\avgop}[1]{\left\langle {\protect #1} \right\rangle}

\makeindex

\begin{document}

\title*{Machine Learning approaches to classical density functional theory}
\author{Alessandro Simon and Martin Oettel}
\institute{Alessandro Simon \at Institute of Applied Physics, University of T{\"u}bingen, Germany, \email{alessandro-rodolfo.simon@uni-tuebingen.de} \and Martin Oettel \at Institute of Applied Physics, University of T{\"u}bingen, Germany, \email{martin.oettel@uni-tuebingen.de} }

\maketitle

\large{
  The following chapter is set to appear in:\newline te Vrugt, M. (Ed.), (2024), \emph{Artificial Intelligence and Intelligent Matter}, published by Springer (Cham).
}

\tableofcontents
\markboth{}{}

\abstract{
In this chapter, we discuss recent advances and new opportunities through methods of machine learning 
for the field of classical density functional theory, dealing
with the equilibrium properties of thermal nano-- and microparticle systems having classical 
interactions. Machine learning methods offer the great potential to construct and/or improve the free energy functional
(the central object of density functional theory) from simulation data and thus they complement traditional
physics-- or intuition--based approaches to the free energy construction. We also give an outlook to machine learning
efforts in related fields, such as liquid state theory, electron  density functional theory and power functional
theory as a functionally
formulated approach to classical nonequilibrium systems. 
}

\section{Introduction}

{
Density functional theory (DFT) is a powerful reductionist scheme for classical and quantum many-body systems in equilibrium. The reduction comes about by the existence of a unique (free) energy functional, depending only on the one--body density of classical or quantum particles. From this functional, all other properties of interest, most notably higher order correlations can be derived.     
For quantum systems at zero temperature, $T=0$, the unique mapping between an external potential $\vext(\rv)$ and the particle density $\rho(\rv)$ entails the existence of a unique energy functional $E[\rho]$, not depending on $\vext$ \cite{hohenberg1964inhomogeneous}. For finite $T$, the argument can be generalized to show the existence of a unique free energy functional $\FE[\rho]$ both in the quantum case \cite{mermin1965thermal} and in the classical case \cite{evans1979nature}. However, in general the functional $\FE[\rho]$ is different for differing internal Hamiltonians of the system (i.e. differing particle--particle interaction potentials) and it is not known except for a few exceptional cases (like the ideal gas and a system of one--dimensional (1D) hard rods in the classical case). Moreover, in the classical case there are a number of Hamiltonians of interest, ranging from those of atomic and molecular systems to those of polymeric and colloidal systems where the basic particles are macromolecular in nature and their particle--particle interactions are already coarse--grained. 
}

{
Constructing classical free energy functionals is occasionally more of an art than a systematic procedure and entails the use of specific physical and mathematical insight into the system of interest. For hard--body systems, e.g., one can use concepts from integral geometry to derive fundamental  measure theory (FMT) \cite{roth2010fundamental}. These FMT functionals are perhaps examples for being most advanced and accurate compared to simulation data (however, they are not exact except for the 1D case). For systems with other interactions one has not reached yet such a level of insight and precision. Here, efforts have gone into defining simplified model systems for which functionals are constructed with differing success. These simplified model systems include simple fluids with repulsive cores and attractive tails (as embodied by the Lennard-Jones (LJ) potential) \cite{lutsko2008}, particles with (screened) electrostatic interactions \cite{bueltmann2022}, polymeric fluids with simple connectivity assumptions between monomers \cite{forsman2017}, patchy particles as examples for associating fluids \cite{stopper2018bulk}, .... An exception to these simplified model systems is water (due to its overwhelming importance) whose functional building can be viewed as archetypical for liquids with anisotropic molecules \cite{ding2017efficient}.  
}

{
The ``cheap'' alternative to the difficulties of classical DFT (cDFT) are classical simulations, either Monte Carlo (MC) or Molecular Dynamics (MD). One simply needs to specify the underlying potential energies and forces and acquires the desired properties as statistical averages over snapshots of the system. Such simulations, however, can be costly if higher--order correlations are needed, or if free energies need to be computed via thermodynamic integrations. Additionally, physical insights into the system (such as schematic behavior of certain correlations) necessitate running simulations for a large number of parameters (thermodynamic ones such as density and temperature, or specific parameters in the interaction Hamiltonian). Nevertheless simulation data constitute ``ground truth'' for a classical model which, as said, can be generated in a relatively cheap way, and this situation appears to be highly suitable for big data techniques as exemplified by machine learning (ML). Thus, there is hope to combine the precision of simulation data with the conceptual power of the cDFT formulation, which in the end also would allow for very resource--efficient computations. A vision for the description of a classical many--body system would consequently be the systematic use of simulation data to construct or ``learn'' an interpretable and  manipulable (functionally differentiable) free energy functional. The systematic use would include the possibility to refine and improve the ``learned'' functional if needed. 
}

{
This small review intends to cover the efforts of the past years to apply ML techniques to cDFT. In Sec.~\ref{sec:cdft}, we briefly summarize basic relations of density functional theory. Work in the past years has concentrated on the ``simplest'' of the simplified model systems mentioned above, and these are introduced in Sec.~\ref{sec:modelsystems}. Specific approaches are reviewed in Sec.~\ref{sec:approaches}. At the moment, there appears to be no preference for or clear advantage of a specific ML technique, so we attempt to describe the gist of the used techniques in a tractable manner in this section. Finally, in Sec.~\ref{sec:quantum_pft} we describe the relation to the integral equation method of liquid state theory, to some ML approaches to the problem of electron (quantum) DFT and give an outlook to the general nonequilibrium, time--dependent problem which allows a formulation akin to cDFT in terms of a unique power functional \cite{schmidt2013power}.   
}

\section{Classical DFT: basic theory}
\label{sec:cdft}

There are excellent books on classical liquid state theory and more specifically excellent reviews on cDFT, for a selection we refer to
\cite{evans1979nature, roth2010fundamental, hansen2013theory, solana2013perturbation}.

We consider rigid particles with positional and orientational degrees of freedom, thus particles can be anisotropic to allow for the description of molecular fluids and nonspherical colloidal systems.   
We follow standard classical statistical mechanics in the grand canonical ensemble. For the Hamiltonian we assume the following form
\begin{equation}
    \label{eq:def_hamil}
    H = K + u(\rv^N, \omega^N) + \sum_i \vext(\rv_i, \omega_i)
\end{equation}
where $K$ is the kinetic energy of translational and rotational motion. Furthermore,
$\rv_i$ is the position, 
and $\omega_i$ the orientation (in general specified by three Euler angles) of particle $i$. The position--dependent part of the internal energy $u(\rv^N, \omega^N)$ is often taken to be a sum of 2--body pair potentials but this is not necessary. The external potential $V^\rmext(\rv_i, \omega_i)$ is a one--body term acting in general on both position and orientation of the individual particle. We introduce the collective variable $x_i = [\rv_i, \omega_i]$ combining both position and orientation for brevity. The one-body density is then defined as the statistical average of all particles' positions and orientations
\begin{equation}
    \label{eq:def_onebody}
    \rho(x) =  \avg{\sum_i \delta(x - x_i)}
  \end{equation}
where $\delta$ is the (Dirac) delta function.
Classical density functional theory is based on the existence of a functional for the grand potential $\Omega[\rho(x)]$ whose minimization gives the equilibrium density $\req(x)$. The functional $\Omega[\rho(x)]$ reads 
\begin{equation}
    \label{eq:grand_pot}
\Omega[\rho] = \FE_\rmid[\rho] + \FE_\rmex[\rho] + \int \odif{x} \left (\vext(x) - \mu \right)
\end{equation}
Here, the one--body piece containing $\vext$ and the chemical potential $\mu$ is separated out, and $\FE[\rho]=\FE_\rmid[\rho] + \FE_\rmex[\rho]$ is the unique free energy functional only depending on the density (and not on the external potential).
It consists of the ideal (non-interacting) part $\FE_\mathrm{id}[\rho]$ and the excess (over ideal) part $\FE_\rmex[\rho]$.
The ideal gas part
is given by
\begin{equation}
  \label{eq:fid}
   \beta \FE_{\text{id}} = \int \odif{x} \, \rho(x) \left[\ln(\rho(x) \lambda^{3}) -1  \right] \,,
\end{equation}  
and is the exact free energy functional for noninteracting particles ($u=0$). Here, $\beta=1/(\kb T)$ is the inverse temperature and $\lambda^3$ is a volume factor containing the de--Broglie thermal wavelength and a normalization factor of the orientational integral.
Minimization of $\Omega$ w.r.t.\ the density results in the Euler--Lagrange equation
\begin{equation}
    \label{eq:el}
    \req(x) = \exp \left( - \beta \vext(x) + \beta \mu + c_1 [\req(x)]) \right)
\end{equation}
Here, $c_1[\rho]$ is the first member of the hierarchy of direct correlation functions (DCF), defined by functional derivatives of $\beta\FE_\rmex[\rho]$ w.r.t. the density. Specifically,
\begin{equation}
  c_1(x)[\rho] = - \beta  \frac{\delta \FE_\rmex[\rho] }{\delta \rho(x)} \;,
\end{equation}
and (owing to its importance)
\begin{equation}
  c_2(x,x')[\rho] = - \beta  \frac{\delta^2 \FE_\rmex[\rho] }{\delta \rho(x) \delta\rho(x')} \;,
\end{equation}
is the second--order direct correlation function (often ``the'' DCF in the literature).

The excess free energy functional $\FE_\rmex[\rho]$ is in general not known. The most famous exception is the system of 1D hard rods which therefore has played an important role in the past years to test ML methods (see below). The full functional $\FE_\rmex[\rho]$ is not directly accessible in simulations (as ``ground truth''), easily computable is only the equilibrium density profile $\req(x)$ for a chosen $\vext(x)$. The Euler--Lagrange equation (\ref{eq:el}) entails a map
\begin{equation}
   \req(x) \quad \leftrightarrow \quad  c_1(x) [\req(x)]\;,
\end{equation}
and thus simulations can provides us with individual points $\{\req,c_1\}$ of this map.
Thus the reconstruction of the functional $c_1(x) [\req(x)]$ should be  
suitable for ML methods given enough points of the map. Having learned the functional $c_1(x) [\req(x)]$ gives access to higher--order correlation functions and specific physics contained in those (e.g. sum rules) as long as the ML methods allow for functional differentiation.  

Note that the original DFT proof \cite{mermin1965thermal, evans1979nature} rests on the unique map
\begin{equation}
   \req(x) \quad \leftrightarrow \quad  \vext(x) \;.
\end{equation}
If the functional of the excess free energy is not known, the arrow to the right from $\req$ to $\vext$ is actually a typical difficult inverse problem for simulations. Given that simulations allow the computation of $\req(x)$ for given $\vext(x)$ with comparable ease (arrow to the left), ML methods should be in principle suited to learn the functional $\req(x)[\vext(x)]$ which from the functional perspective complements $c_1(x) [\req(x)]$ as follows. We define a ``local chemical potential'' by $\psi(x)=\mu-\vext(x)$. The grand potential functional $\Omega$ of Eq.~\eqref{eq:grand_pot} can be viewed as a functional of $\psi(x)$ whose functional derivatives generate the density profile and the higher--order density fluctuation functions \cite{evans1979nature}. Specifically
\begin{equation}
  \frac{\delta \Omega[\psi]}{\delta\psi(x)} = - \req(x)
\end{equation}
and 
\begin{equation}
  \frac{1}{\beta}\frac{\delta^2 \Omega[\psi]}{\delta\psi(x)\delta\psi(x')} = - H_2(x,x')
\end{equation}
where $H_2(x,x')$ is the density--density correlation function defined by
\begin{equation}
H_2(x,x')=  \avgop{ \left[ \rho(x) -\avgop{ \rho(x) } \right] \left[ \rho(x')- \avgop{ \rho(x')} \right]}
\end{equation}
One can write $H_2(x,x')=\req(x)\delta(x-x') + \req(x)\req(x')h(x,x') $ where $h(x,x')$ is linked to the standard pair correlation function $g(x,x')$ by $h=g-1$. The functions $h(x,x')$ and $c_2(x,x')$ are linked by the famous Ornstein--Zernike relation
\begin{equation}
 \label{eq:oz}
  h(x,x') - c_2(x,x') = \int dx'' h(x,x'') \, \req(x'')\, c_2(x',x'') \;,
\end{equation}
an integral equation of formidable difficulty in the case of anisotropic fluids and a general $\vext(x)$.

The link between the structural functions $h$ and $c_2$ is the problem of the integral equation approach to liquid state theory and can be seen as a specific subtopic of the general cDFT problem. We will comment upon recent ML advances in integral equation theory briefly in Sec.~\ref{sec:ie}.

\section{Model systems}
\label{sec:modelsystems}
The ML methods which are described more in detail below in Sec.~\ref{sec:approaches} concentrate on different aspects of cDFT and also apply to different model systems. Here, for completeness, we briefly introduce the used model systems. 

\subsection{Hard sphere system in 1D and 3D}
\begin{figure}
    \centering
    \includegraphics[width=.7\linewidth]{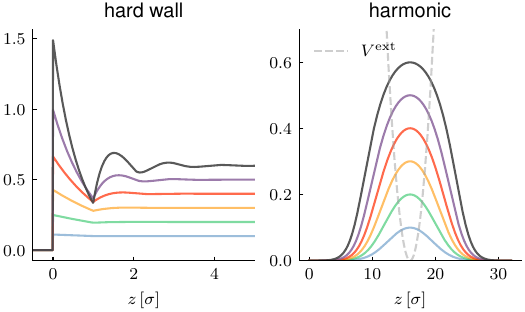}
    \caption{Solutions for the density profiles $\rho(z)$ of the one dimensional hard-rod system for two kinds of external potentials. The bulk densities are linearly increasing $\rho_b \sigma = 0.1, 0.2, \ldots, 0.6$, and
    the rod length is $\sigma$.
    }
    \label{fig:demo-dft}
\end{figure}

An easy to handle, yet non-trivial model is the one dimensional hard-rod system. Here the particles of width $\sigma$ are constrained to a line with coordinate $z$
without the possibility of overlapping and no additional interaction between them.
It is also one of the few models where the exact excess free energy functional is known.
\begin{equation}
    \label{eq:1dhr_fe}
    \beta \Fex^\text{hr}[\rho(z)] = \int \Phi(n_0, n_1) \odif{z} = - \int n_0 \ln (1 - n_1) \odif{z}
\end{equation}
with the weighted densities 
\begin{equation}
    n_i(z) = \int \odif{z'} \rho(z')  \omega_i(z-z') = \rho \otimes \omega_i
\end{equation}
and the two kernel functions $\omega_0(z) = \delta(\sigma/2 - |z|)/2$ and 
$\omega_1(z) = \Theta(\sigma/2 - |z|)$.
This exact functional has a form  characteristic for many of the approximate functionals. The local excess free energy density, $\beta \fex(z)=\Phi(z)$, depends non-locally on the density profile $\rho(z)$ through weighted (or ``smeared'') densities with characteristic weight functions. 
In Fig.~\ref{fig:demo-dft}, we show characteristic density profiles resulting from this exact functional, namely adsorption at a hard wall (left panel, showing the characteristic layering effect) and confinement in a trapping potential (right panel). Such profiles are frequently used in the machine learning routines described later.

The hard rod system can also be extended to higher dimensions, although no exact
functionals are known for the 2D or 3D case. There are however very accurate 
functionals based on fundamental measure theory (FMT), see Ref.~\cite{roth2010fundamental} for a review.

\subsection{Lennard-Jones}
The Lennard-Jones (LJ) system is undoubtedly one of the most extensively studied models for a simple fluid, featuring typical short--range repulsion (Pauli exclusion of closed electron shells) and longer--ranged attraction (van der Waals interaction), giving rise to a gas, liquid and solid phase. It is realistic for noble gases. The LJ potential has the form
\begin{equation}
    u_\mathrm{LJ}(r) = -4 \varepsilon \left[ (r/\sigma)^6 - (r/\sigma)^{12} \right] \;,
\end{equation}
where $\sigma$ is a particle diameter.
No exact free energy functional for the LJ system is known, owing mainly to the complications of the attractive part. Here, the random phase approximation~\cite{hansen2013theory,solana2013perturbation} (RPA) is semiquantitative for supercritical state points at high temperature but gives also insights and a qualitative account of the phase diagram for lower temperatures. The RPA functional is given by:
\begin{equation}
    \label{eq:rpa}
    \Fex^\text{RPA}[\rho] = \Fex^\text{ref}[\rho] + \frac{1}{2} \int \odif{\rv} \odif{\rv'} \rho(\rv) \rho(\rv') u_\mathrm{LJ}^\text{att}(\rv-\rv')
\end{equation}
where $\Fex^\text{ref}$ is a reference functional (usually the one of a hard sphere system with optimized $\sigma$).
The attractive part is of typical mean--field form (with the assumption of uncorrelated densities), and through
defining $n^\text{att}=\rho \otimes u_\mathrm{LJ}^\text{att} $ can be written in the weighted density form described above. $u_\mathrm{LJ}^\text{att}$ is a suitably defined attractive part of the LJ potential, e.g. from the WCA prescription \cite{hansen2013theory,solana2013perturbation}. 

\subsection{Kern--Frenkel}

\begin{figure}
    \centering
    \includegraphics[width=.5\linewidth]{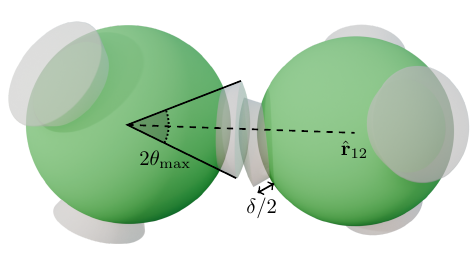}
    \caption{Patch--patch interaction in the Kern--Frenkel model between two particles connected by the vector $\hat{\rv}_{12}$. The angle $\theta^\mathrm{max}$ specifies the opening of the patch cone and the parameter
     $\delta$ the extension in the radial direction.}
    \label{fig:kf}
\end{figure}

The Kern--Frenkel (KF) potential is a popular model for anisotropic interactions. In addition to the repulsive hard-sphere interaction, every particle is equipped
with $N$ so called patches, which in the KF model are cones emanating from the particle center outwards up to a certain cut-off radius. If any two cones belonging to different particles overlap, they are considered to be bonded, with an associated decrease of the energy of the system by $\varepsilon$ (see Fig.~\ref{fig:kf}). The bonding energy, the cut-off radius $\delta$, together with the cone angle $\theta$ and number of patches are tunable parameters
which allow us to adapt the model to different situations. 
Popular are choices for which ensure that the single bond condition between two particles is fulfilled, which is the particular limit of an associating fluid where Wertheim theory \cite{solana2013perturbation} gives a semiquantitative account of the phase diagram and bonding statistics.

\section{Machine learning approaches}
\label{sec:approaches}

The machine learning approaches in the literature can be roughly classified into the following categories: 
\begin{itemize}
    \item Parameterization of the excess functional \cite{shang2019classical, shang2020FEQL, cats2021, simon2024machine, dijkman2024learning} ($\req \rightarrow \FE$ map)
    \item Parameterization of the one-body correlation function \cite{sammuller2023neural, sammuller2023neural_ii} ($\req \rightarrow c_1$ map)
    \item (Parametric) Bayesian methods on the $\req \leftrightarrow \vext$ map \cite{yatsyshin2021,malpica2023physics}
    \item Gaussian Processes for the $\req \rightarrow \vext$ map for reinforcement learning \cite{fang2022reliable}
\end{itemize}

Below we try to categorize the approaches according to their main idea/ingredient, even though one should be aware of possible overlaps.

\begin{table}[h]
\centering
\begin{tabularx}{\textwidth}{@{}XllXl@{}}
    \toprule
    Approach & Input & Output & Parameters & Loss \\
    \midrule
    Direct parameterization \cite{shang2019classical,shang2020FEQL,cats2021, simon2024machine} & $\mu, V^\rmext(\rv), (\beta)$ & $\FE_\mathrm{ex}[\rho]$ & interaction kernels, symbolic structure  & $\MSE (\rho^\star, \rho)$  + reg.\\
    Neural DFT \cite{sammuller2023neural,sammuller2023neural_ii} & $\rho(\rv)$ & $c_1(\rv, [\rho])$ & network weights  & $\MSE (c_1^\star, c_1)$ \\
	$c_2$-matching~\cite{dijkman2024learning} & $\rho(\mathbf{z})$ & $c_2(|z_i-z_j|)$ & networks weights & $\MSE (c_2^\star, c_2)$ + reg.\\
    Bayesian model~\cite{malpica2023physics} & $\rho(\rv)$ & $V^\rmext(\rv)$ & external potential parameterization & [Posterior sampling] \\
    Gaussian process~\cite{fang2022reliable} & $\mu, V^\rmext(\rv)$ & $\rho(\rv)$ & prior hyperparameters & [Posterior sampling]\\
 
    \bottomrule
\end{tabularx}
\caption{Different methods and their respective input/output pairs and loss terms, if applicable. The star superscript denotes the outputs of the neural network.}
\label{tab:methods_tab}
\end{table}

\subsection{Direct parameterization of the functional}
\label{sec:direct-param}

In the historical development of cDFT for simple fluids, it was quickly noticed that the local density approximation
\begin{equation}
    \Fex[\rho] = \int \odif{\rv} \, \fex(\rho(\rv)) \, ,
\end{equation}
or the (square) gradient approximation
\begin{equation}
    \Fex[\rho] = \int \odif{\rv}  \, \left[  \fex(\rho(\rv))  + \frac{a}{2} \left( \nabla \rho(\rv) \right)^2  \right ] 
\end{equation}
have very limited accuracy and cannot capture especially the correlation effect due to repulsive cores (``layering''). Nevertheless, the square gradient approximation contributed a lot to our understanding of  the physics of liquid--vapour interfaces.
For the subsequent development, the insight was crucial that the functional has a non-local
dependency on the density distribution $\rho(x)$ through convoluted, weighted densities
$ n_i =\rho \otimes \omega_i$ with weight functions $\omega_i$ which are usually of finite range, see also Sec.~\ref{sec:modelsystems}.
The weighted-density form for the free energy entails that the minimizing equation for the equilibrium profile $\req$, Eq.~(\ref{eq:el}), is a nonlinear integral equation.  
It is usually solved using the Picard iteration scheme, 
which starts with an initial guess $\rho^{i=0}(\rv)$, usually the uniform bulk density.
Inserting $\rho^i$ into the EL equation gives a new density distribution
\begin{equation}
  \label{eq:it1}
    \underline{\rho^{i}}(\rv) = \exp \left( -\beta V^\rmext(\rv) + \beta \mu + c_1[\rho^{i}(\rv)] \right) \,.
\end{equation}
Owing to stability reasons, the next iteration $\rho^{i+1}$ is obtained by mixing: 
\begin{eqnarray}
   \label{eq:it2}
    \rho^{i+1}(\rv) = (1-\alpha) \rho^{i} (\rv) +  \alpha \underline{\rho^{i}}(\rv) \;,
\end{eqnarray}
where $\alpha$ is a small parameter which can be dynamically adapted. Upon reaching a small norm of the difference between two iterations, $||\rho^{i+1}(\rv), \rho^{i}(\rv) || \leq \epsilon$, one can speak of a ``self--consistent'' solution, i.e. the free energy functional permits a minimization of the grand potential functional up to a certain numerical  accuracy. For machine--learned functionals, this is a nontrivial condition.

In the context of machine learning, it appears natural to assume (parametrize) a new excess functional $\Fex^\mlrm[\rho; \theta]$, which depends on some unknown internal parameters $\theta$. These parameters could either constitute the weights of a universal approximator (e.g.\ a multilayer perceptron (MLP)) or the variables in an \textit{ansatz} built on existing knowledge of the system. 
Using a black-box model such as an MLP, prevents us from any direct interpretation of the learned internal representation. On the other hand, less general, parameterized models might suffer from limited generalizability. 
One may roughly differentiate between the two (idealized) camps of ML/DFT practitioners:\\
(i) The ones mainly interested in an accurate emulation of the physical system in question, taking advantage of the efficiency of the DFT formalism and \\ 
(ii) those interested in uncovering (``fitting'') interpretable representations of the functional maps.

\subsubsection{Mean-field and third order terms, isotropic case}
\label{sec:mfLJ}

The LJ system was investigated in 1D \cite{shang2019classical} and then later in 3D \cite{cats2021} using a ``camp (ii)'' approach aiming at learning mean-field and higher order correction terms for the attractive part of the interaction.   
Apart from one important detail, namely whether self-consistency of the functional was imposed during training,
both approaches are similar and we will limit the exposition to the more recent Ref.~\cite{cats2021}. The topic of self-consistency will be brought up again at the end of this section.

In the work by Cats et al.~\cite{cats2021} the authors consider the LJ fluid in the reference scheme. The standard RPA functional of Eq.~(\ref{eq:rpa}) is compared with the RPA functional plus ML corrections, where the corrections are parametrized as
\begin{align}
\beta \Delta \Fex^{\mathrm{ML2}} &=  \frac{1}{2} \int \odif{\rv}\odif{\rv'} \rho(\rv)\rho(\rv') \Omega_2(|\rv-\rv'|) 
\label{eq:cats2}\\
\beta \Delta \Fex^{\mathrm{ML3}} &=  \frac{1}{3} \int \odif{\rv}\odif{\rv'} \rho(\rv)^2\rho(\rv') \Omega_3(|\rv-\rv'|) \label{eq:cats3}
\end{align}
with the unknown kernels $\Omega_i$. 
Here, ML2 is second order in density and provides a correction to the RPA mean--field kernel, while ML3 adds a contribution of third order in density.
Training was performed at one supercritical temperature in flat wall geometry with external potentials (varying in their steepness), here $\rho(z)$ only depends on the Cartesian coordinate $z$. In the flat wall geometry, the functionals (\ref{eq:cats2},\ref{eq:cats3}) retain their form  with $\rv,\rv' \to z,z'$ and $\Omega_i \to \omega_i(z)$ and the new kernels $\omega_i(z)$ are differentiable parameters of the network. Owing to the isotropy of the LJ interaction, the kernels are related by   
\begin{align}
    \omega_i(z) &= 2 \pi \int^\infty_z \odif{r} r \Omega_i(r) \label{eq:omega_rtoz} \;,\\
    \Omega_i(|\rv|) &= - \frac{1}{2\pi} \frac{1}{z} \odv{\omega_i(z)}{z}_{z=|\rv|} \label{eq:omtoOm}\;,
\end{align}
thus the training in the flat wall geometry is sufficient, and the resulting functional can be used in any other geometry.

\subparagraph{Training}

The loss consists of two parts. The principal term $L_1$ quantifies the difference between the profiles resulting from the parameterized ansatz
and the Monte Carlo data. The regularizer term $L_2$ constrains the interaction kernels to be localized around the center regions and smoothly decaying.
(Regularizers are usually needed to prevent overfitting in ML models.)
In order to evaluate the derivative of the complete loss with respect to the network parameters it is necessary to solve the Euler--Lagrange equation to obtain the ML profile 
$\rho^\star$. 
All other necessary derivatives can be computed analytically once the equilibrium
density distribution is known for a certain set of parameters.

\subparagraph{Results}

The density profiles generated with the learned correction terms show a large improvement compared to the RPA ansatz alone, even for the external potential being highly irregular and very different from potentials used during training. ML3 does not give substantial improvements over ML2 such that the obtained functional is a vindication of the RPA \textit{ansatz} with an optimized kernel (at least for the particular temperature chosen). Here, an extension to a larger temperature region is clearly of interest. (Note that the 1D investigations of Ref.~\cite{shang2019classical} included more temperatures and here a correction to the functional of third order in density showed more substantial improvements.) Results in radial geometry (not the training geometry!) were obtained for the direct correlation function $c_2$ and the pair correlation function and showed good agreement with simulations.

\subparagraph{Self-consistency}

As mentioned earlier, it is possible to either impose self-consistency already during training or to check for it after training is complete.
While it is of course desirable to have this property included into the training already it is not straightforward to do so if the ML functional is somewhat ``noisy''.
The self-consistency condition can only be ensured after solving the Euler--Lagrange equation \cref{eq:el} for $\req$, which may
be time consuming. Further, the solution is usually found using iterative methods which makes the computational graph leading to the solution
grow very fast, posing a problem to automatic differentiation approaches.
In ref.~\cite{shang2019classical} this issue was circumvented by using only the generative output of 
the rhs of the Euler--Lagrange equation (\ref{eq:el}) as $\rho^\star$, using the ML functional but evaluated with the ground truth $\rho_\text{sim}\approx \req$ from simulations. 
Having $\rho^\star \approx \rho_\text{sim}$ up to a reasonable precision is a necessary condition for the existence of the fixed point but not a sufficient one.
The approach used in ref.~\cite{cats2021} is based on manually calculating the necessary partial derivatives in order to do a back-propagation pass
with respect to the parameters of the network. However, more complicated \textit{ans\"atze} will make this procedure more laborious.

\subsubsection{A mean-field functional for the Kern--Frenkel fluid}
\label{sec:kf}

Similar in method (reference functional plus ML mean--field functional) is the investigation of an anisotropic fluid in Ref.~\cite{simon2024machine}.
A major complication results from the fact that the model fluid interacts through the angle-dependent 
Kern--Frenkel potential, making it necessary to include orientational degrees of freedom beyond the orientationally averaged density distribution alone. The training was performed on simulated density profiles between hard walls for a range of densities and supercritical temperatures not far from the critical point. In this geometry, the density profile $\rho(x) \to \rho(z) \alpha(z,\omega)$ is the product of an orientationally averaged profile $\rho$ and a position-- and orientation--dependent orientational profile $\alpha$.

The approach identifies a set of reduced orientational profiles $\alpha_i(z)$ invariant under
the symmetry group of the particle (tetrahedral symmetry) and include all linearly-dependent combinations.
The excess free energy $\fex=\fex^\text{ref}+\fex^\mlrm$ (here per area) is then augmented by the ML mean-field term
\begin{equation}
    \beta \fex^\mlrm[\rho, \{\alpha_i\}] = \frac{1}{2} \int \odif{z} \odif{z'} \rho(z) \rho(z')\sum_{ij} M^{ij}(z-z') \alpha^i(z) \alpha^j(z')
\end{equation}
where the kernels $M^{ij}(z)$, discretized on a grid, are the differentiable parameters of the network. The reference part comes from ``functionalized'' Wertheim theory for associating fluids and depends on the orientationally averaged profile $\rho$ only. 

\begin{figure}
    \centering
    \includegraphics[width=.6\linewidth]{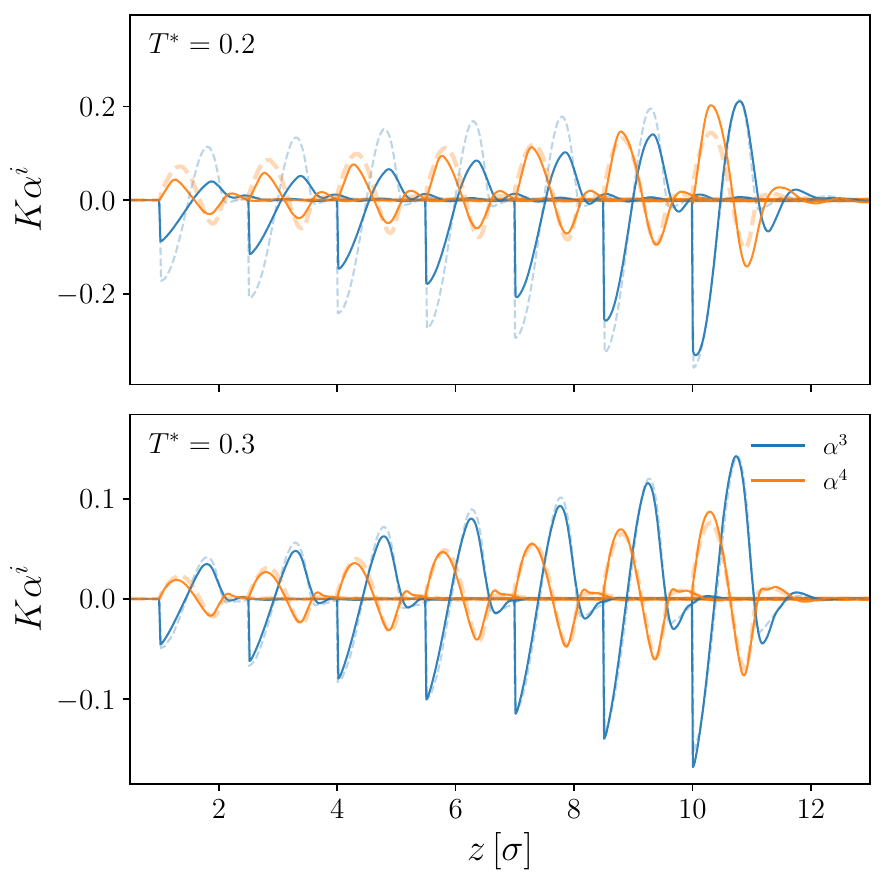}
    \caption{Self-consistent, leading orientational moments from learned mean-field ansatz for two isotherms with reduced temperature $T^*=0.2$ and 0.3. The $y$ axis shows the first two (scaled) orientational moments at distance $z$ to the wall, which starts at $z=0.5$ in comparison to simulations (faint dashed lines). The result for each bulk density (in the range $\rho\sigma^3$ between $0.1$ and $0.3$, covering the critical region) was shifted to the right by $\Delta z = 1.5$, starting from the lowest density on the far left. The critical temperature is $T^*\approx 0.17$.
    Reprinted with permission from Ref.~\cite{simon2024machine}. 
    }
    \label{fig:ori-norho}
\end{figure}

Due to the larger state space, the training data (especially for the orientations) is more noisy than that of an
isotropic systems. This makes training by evaluating on the fixed point alone difficult, as it introduces unphysical
components to the learned parameters. It is therefore necessary to evaluate the derivatives of the loss with respect to a self-consistent solution of the Euler--Lagrange equations (fixed point), similar to ref~\cite{cats2021}.
While this approach is already sufficient to constrain the fixed point in a numerically stable way, it is 
sensible to assume that the gradient for points in the vicinity of the fixed point can help to stabilize the
procedure further. This could be done, for example, by saving the computational graph of the whole iterative procedure up to the final fixed point, 
in order to later do a backpropagation on it. Unfortunately this becomes too memory intensive even for relatively few iterations (say in the hundreds). There are some techniques to save memory on repeated function evaluations in loops (e.g.\ \texttt{jax.lax.scan}) and also more sophisticated solvers that can reach a solution after fewer iterations. Another way to reduce memory consumption is by using implicit differentiation. One trades, effectively, the memory savings
for the need of solving a linear equation for every backpropagation. This can however be done, rather fast.
For a fixed point Euler--Lagrange equation of the form $\alpha=g(\alpha,\theta)$ (both $\alpha,\theta$ are vectors) one finds the gradient of the fixed point $\alpha^\star$
\begin{equation}
    \pdv{\alpha^\star}{\theta} = \pdv{g}{\theta}_{\alpha^\star} \left[ \mathbb{I} - \pdv{g}{\alpha}_{\alpha^\star} \right]^{-1}
\end{equation}
While this result is exact, solving the matrix inversion can become unstable.
By approximating the inverse with just the identity matrix, one can save computing time and
in many cases stabilize the procedure \cite{chang2022object}.

\subparagraph{Results}

The mean field kernels were trained on different isotherms since there is an implicit temperature dependence of the kernels, that is not explicitly
included. As expected from a mean-field ansatz, the reproduction of the observed orientation works better for high-temperature state points.
At lower temperatures, but still above the critical temperature, the mean-field ansatz is not able to accurately represent the orientations
for all considered bulk densities (average reduced densities $\rho\sigma^3$ ranged from \num{0.10} to \num{0.30}), see Fig.~\ref{fig:ori-norho}.
In every case, the trained kernels were
much stronger than those predicted by RPA, which were computed using Monte Carlo integration of the relevant orientational integrals. This is very different to the LJ case.

Due to the training method that was used, the functional could be minimized self-consistently
and could be extrapolated to higher densities. As opposed to the case of an isotropic fluid, the ML functional with the learned kernel $M^{ij}$ can not be used for other geometries since in the flat wall geometry the elements of a general mean-field kernel are projected and integrated, and can not be reconstructed.

\subsubsection{Equation Learner Network}

\label{sec:eql}
The approach used in ref.~\cite{shang2020FEQL} 
makes use of the Equation Learner Network (EQL) \cite{martius2016extrapolation, sahoo2018learning}, 
which is a neural network used for symbolic regression, i.e.\ the problem of finding symbolic expressions that 
describe the relationship between two datasets.
The goal in this context was to discover the symbolic form 
the free energy functional and not just an opaque representation of it. When using the EQL it is necessary to specify beforehand the kind of building blocks 
(\textsf{sin, cos}, $\times, \div$ etc.) that can appear in the learned
expression. By additionally specifying the maximum depth or number of layers, the amount of representable expression is fixed and finite. It is therefore not guaranteed that the EQL is able to find the exact equations with the selected basis functions, but the interpretability of the results might be able to steer further
investigations.

The main advantage of the EQL consists in being fully differentiable during training, one can therefore 
not only use its output $f(x)$ to compute losses (and their gradients, w.r.t.\ $\theta$) but also
$f'(x), f''(x)$ and so on. Indeed this is needed for the evaluation of  
$c_1=-\beta \delta \Fex /\delta \rho$ during training.%

\begin{figure}
    \centering
    \includegraphics[width=.7\linewidth]{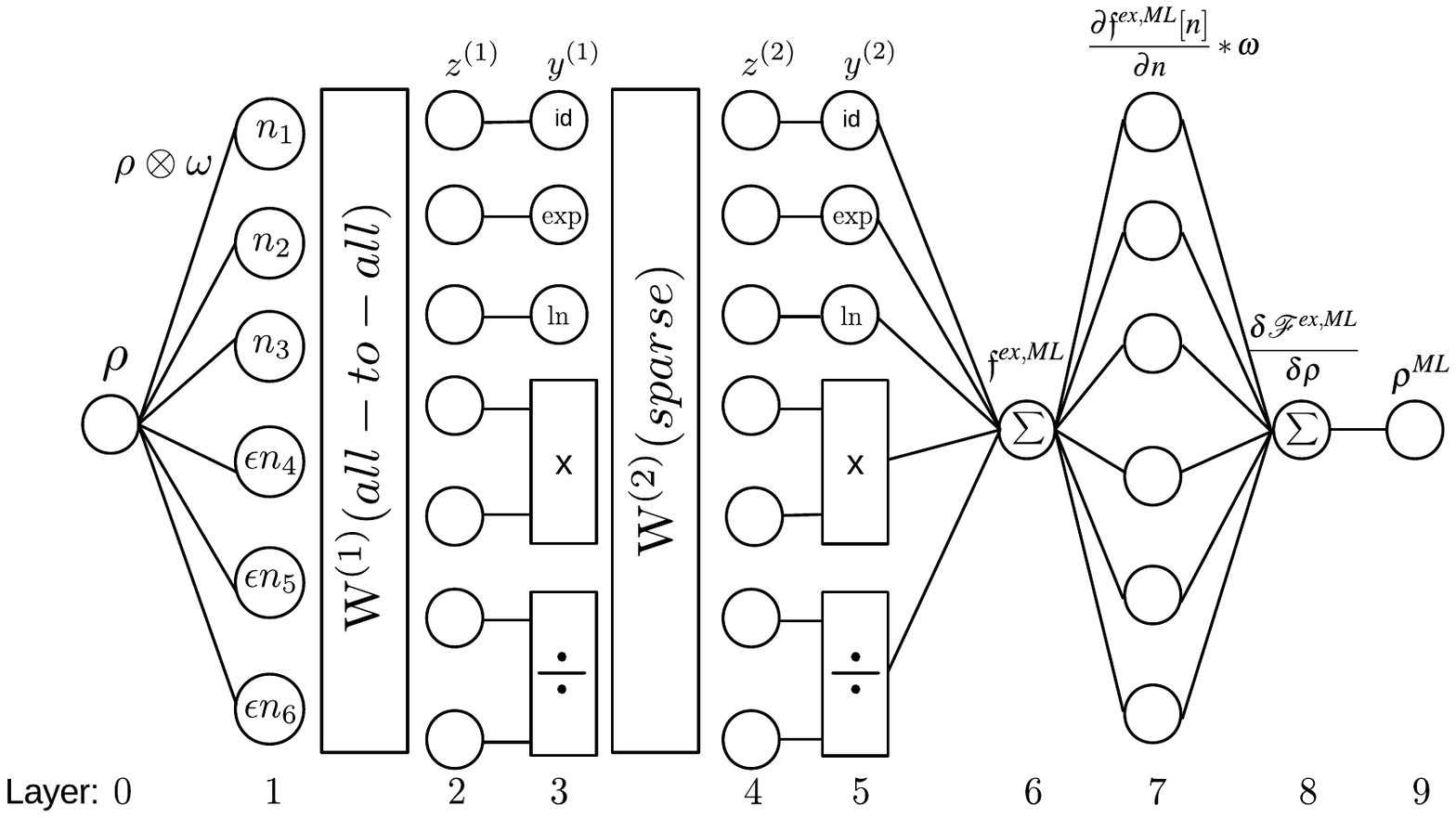}
    \caption{Architecture of the functional equation learner network for learning $\Fex$ for the 1D LJ system. The density is first fed into a convolution layer with learnable kernels, defining six weighted densities $n_i$. Three of the $n_i$ are multiplied with the reduced LJ interaction strength $\epsilon=\beta\varepsilon$ in order to capture the temperature dependence. After that, two layers
    of non--linear symbolic transformations follow, producing finally the ML free energy density (here denoted as $f^{ex,ML}$). The discretized functional derivative is performed and the resulting ML density profile produced, see Eq.~(\ref{eq:rhostarSam}). Reprinted from~\cite{shang2020FEQL}, with the permission of AIP Publishing.}
    \label{fig:feql}
\end{figure}

The network was applied to the 1D hard rod and 1D LJ--like system and functionally represents the free energy density
$\fex[\rho;\theta]$ where $\theta$ are the parameters of the network.
An example for the network, aiming at the full functional for the LJ system, is shown in Fig.~\ref{fig:feql}.
To train the network on ground truth density distributions $\req(\rv)$, 
the predicted ML output density profile $\rho^\star =\rho_\mlrm$ is defined as in Ref.~\cite{shang2019classical}: 
\begin{equation}
    \rho_\mlrm(\rv) = \rhob \exp \left( - \beta V^\rmext + \beta \mu_\rmex + c_1^\star[\req(\rv)]  \right)
   \label{eq:rhostarSam}
 \end{equation}
where the chemical potential was split into the ideal part $\mu_\mathrm{id} = \kb T \lambda^3 / \rho$ and the excess part $\mu_\rmex$.
For the 1D hard rod fluid, the training data can be generated fast and exact using the functional in Eq.~(\ref{eq:1dhr_fe}), and for the LJ fluid ground truth data were simulated 
using Monte Carlo methods.

\subparagraph{Results}

For the hard rod system, the network in Fig.~\ref{fig:feql} was used with just the upper three weighted densities. Since no other training constraints besides the density profiles were used, the network could not find the exact functional (see the SI of Ref.~\cite{shang2020FEQL}) and also not the exact virial coefficients but the overall agreement for the equation of state and density profiles was very good, also outside the training region.   

For the LJ system, two approaches were pursued. On the one hand, the reference \textit{ansatz} was used in the form  
\begin{equation}
    \label{eq:fex_splitting}
    \Fex[\rho; \theta] = \Fex^\text{hr}[\rho] + \epsilon F_\rmex^\mlrm[\rho; \theta]
\end{equation}
where $\epsilon=\beta\varepsilon$ is the reduced LJ strength and ML accounts for the attraction part of the functional.
Secondly, the full excess functional was learned, using the network in Fig.~\ref{fig:feql}, i.e.\ the temperature dependence (via $\epsilon$) is completely transferred into the ML functional.

The calculated self-consistent density profiles for test and extrapolation showed a good performance.
In addition to the density profiles, the equation of state $P(\rho)$ and direct correlation function $c_2(x)$ was evaluated and compared to the MC case.
In general one sees that the ML version with splitting (``physics informed'')
performs better than the one without, especially for the
direct correlation function, which agrees semiquantitatively 
with simulation.

The complexity of the produced symbolic expressions 
can be controlled by a regularization hyperparameter that penalizes
large values of the expansion parameters (cf.\ $W^{(i)}$ in \cref{fig:feql})
and setting those below a certain threshold (e.g. \num{1e-5}) to zero. 
While being simple, this regularization produces sparsity only indirectly
by putting a pressure on the weights, but without accounting for the 
underlying algebraic structure of the expressions. 
The consequence is, that in order to achieve good results for the
density profiles, a large regularizer value had to be chosen, producing
rather complex expressions which are too complicated indeed
to gain further insight about the mathematical structure of the
functional (see SI of ref.~\cite{shang2020FEQL}).

\subsection{Parameterization of $c_1[\rho]$}
\label{sec:nn-mapping}

\begin{figure}
    \centering
    \includegraphics[width=.5\linewidth]{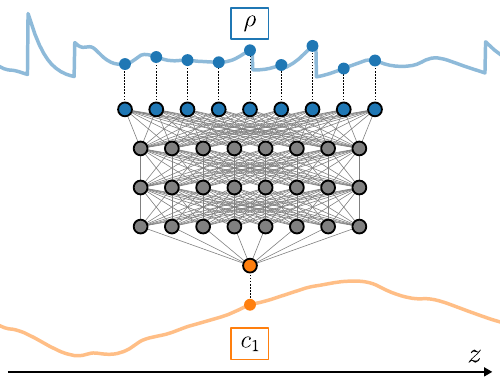}
    \caption{Relationship between input and output of the neural DFT network, parametrized with a standard MLP. The value of $c_1(z)$ depends only on density points within a certain finite interval around $z$.}
    \label{fig:neural_c1}
\end{figure}

Here, the map between the first functional derivative, $c_1=-\beta \delta \Fex/\delta \rho$, and $\rho$ is learned directly \cite{sammuller2023neural,sammuller2023neural_ii} and was dubbed ``neural functional theory'' by the authors. Through Eq.~(\ref{eq:el}), also $c_1$ is obtainable as ground truth from simulations and thus the map can be checked directly on simulation data.
The novelty of this approach is two-fold. 
First, it avoids the use of the self-consistent iteration procedure of Eq.~(\ref{eq:el}) in training by using  the density $\rho(\rv)$ as the input and the one-body direct correlation function $c_1(\rv)$ as the target. An additional helpful observation is that $c_1(\rv)$ is short ranged and thus easier to learn.
The second idea is to use only a part of the input array for inference, instead of the complete information over all of $\rv$. This is physically sensible as the effect of the external potentials cannot extend arbitrarily far in $c_1$ and is mostly felt in the local neighborhood of the point in question, see Fig.~\ref{fig:neural_c1}. Corresponding to this is an imposed finite range of weight functions in the functional equation learner of Fig.~\ref{fig:feql}.

Examples for this neural functional theory were given in Ref.~\cite{sammuller2023neural} for the 3D hard sphere fluid in flat wall geometry and in Ref.~\cite{sammuller2023neural_ii} for the 1D hard rod system. In both cases, density profiles  $\rho(z)$ depend on one Cartesian coordinate. Here we discuss Ref.~\cite{sammuller2023neural}.
The neural network is structured as follows: The input layer consists of \num{513} nodes which are fed with the 
discretized values of the density profile $\rho(z_i)$ for 
    $z_i \in \left[ z-\Delta z, z+\Delta z \right]$ 
i.e.\ only in a window characterized by the cutoff $\Delta z$ around the value $z$.
Then follows a fully-connected multilayer perceptron of three layers and finally
the scalar output $c_1(z)$.
Since $c_1$ is short-ranged, the input window around the point of interest $z$ has enough 
information content for an accurate prediction. Further,
one simulation profile can be split into multiple such windows, which increases the size of the training set considerably. 
A further increase can be achieved by using mirrored profiles.
By making use of automatic differentiation one is able to extract the ``wall geometry'' two-body direct correlation function 
\begin{equation}
    \label{eq:def_c2}
    \bar c_2(z, z'; [\rho]) = \fdv{c_1(z; [\rho])}{\rho(z')}
\end{equation}
which is sufficiently smooth, due to the chosen activation function of the MLP.

\subparagraph{Derived quantities}

Starting from the one-body direction correlation function, which results directly from the network, it is possible to extract other quantities of interest such as the mentioned two-body correlation function.
Note that the system is trained only in the planar wall geometry, but through automatic differentiation (evaluated at bulk densities) 
the bulk 3D direct correlation function $c_2(r)$ is obtainable as $c_2(r) = - \left. \frac{1}{2\pi z} \bar c_2(z)' \right|_{z=r} $, equivalent to the relation between the mean--field kernels (\ref{eq:omtoOm}) of Sec.~\ref{sec:mfLJ}. From that, the structure factor $S(k)$, or the radial distribution function $g(r)$ can be computed using the Ornstein--Zernike relation. Note that the excess free energy is also 
available through functional line integration, which needs to be done numerically
\begin{equation}
    \label{eq:line_integral}
    \beta \Fex[\rho] = - \int_0^1 \odif{\alpha} \int \odif{z} \rho(z) c_1(z; [\alpha \rho])
\end{equation}
The network representation for $c_1$ can be evaluated at any $\rho(z)$, i.e. also along the path.
All these quantities were not included in the original training 
and can therefore be used to reason about the internal consistency of the learned model. 
An additional way to test accuracy and internal consistency is to see whether Noether sum rules (following from symmetry transformations on a thermal system) are fulfilled.

Similar to the case of the anisotropic particles of Sec.~\ref{sec:kf}, the machine--learned $c_1$ is not applicable to problems in other than flat wall geometries. Extensions to genuine 3D situations are certainly desirable.

\subparagraph{Self-consistency}

Since the training process does not involve the self-consistent iteration of Eq.~(\ref{eq:it1}), 
it is not clear a priori, whether the neural $c_1^\star(z)$ can be used to solve the Euler--Lagrange equation
self-consistently and if so, whether the result $\rho^\star(z)$ corresponds to the correct density distribution.
This point was investigated by solving the Euler--Lagrange equation together with $c_1^\star$ using the Picard iteration scheme with mixing.
In order to achieve numerical stability the mixing parameter needed to small in the beginning and could later be increased to a larger value, making convergence faster.
The usual stringent convergence criterion in standard cDFT calculations however needed to be relaxed due to possible fluctuations in the MLP representation $c_1^\star$.

As the training was done on windows around a point of interest, in applications one is not restricted to the box 
sizes that was used for generating the training data. The authors show that the network is able to 
accurately model the density in slits which are larger by an order of magnitude.

\subparagraph{Results}

The model was trained using data from the whole liquid regime with average densities ranging from
$0.003 \sigma^{-3}$ to $0.803 \sigma^{-3}$, and is able reproduce the
one-body correlations up to the accuracy of the input simulation data. This means that the architecture,
with its choice of parameterization and size, is capable of capturing the relevant physics.
Of greater interest is the accuracy of derived quantities like the self-consistent density profiles $\rho(z)$, the DCF $c_2$ and the (integrated) excess free energy. Here we observe that the network is exceeding current standard analytical descriptions, like the FMT excess free energy or the
Carnahan-Starling equation of state, in accuracy and speed. 
This means that in spite of its rather simple design, the neural network is versatile enough to encode
complex many-body information of the hard-sphere fluid beyond the one-body correlation function.

\subsection{Learning $\Fex$ by $c_2$ matching}
\label{sec:c2-matching}

The approach described in Ref.~\cite{dijkman2024learning} is again similar to those mentioned in \ref{sec:direct-param},
as it is based on finding a parameterized, approximative version of $\Fex$, denoted by $F^{(2)}_\theta$. 
This approximative ML network is designed for planar geometry, i.e. $\vext \equiv \vext(z)$ and  
the parameters are not directly interpretable as they constitute the internal representation of 
the neural approximator for the free energy (this is similar to Refs.~\cite{sammuller2023neural,sammuller2023neural_ii}).
As the novel element, instead of using the equilibrium density profile resulting from $F^{(2)}_\theta$ to train the set 
of parameters $\theta$, the authors take
advantage of the full differentiability of the network (here a convolutional neural network or CNN) and calculate
the (planar) DCF by means of automatic differentiation, according to
\begin{equation}
\bar c_2(\theta, |z_i - z_j|) = -\pdv{F^{(2)}_\theta}{\rho(z_i),\rho(z_j)} \frac{ \beta}{A (\Delta z)^2}
\end{equation}
and impose it to be
similar to the (planar) DCF determined from simulations.
The quantity $\Delta z$ is the grid spacing of the discretized grid and $A$
the area of the simulation box in the $xy$ direction.

Initially this training procedure looks like a combination of previous approaches as it is i) learning a parametrized version
of $\Fex$ and ii) using a derivative of $F$ in the loss computation (although a higher-order correlation function than in Ref.~\cite{sammuller2023neural}).
However, using $c_2$ in the loss has a major advantage. Training data of $\bar c_2$ can be extracted in a
rather straightforward way using simulations of homogeneous bulk systems. This is in contrast to
the other approaches, that need as inputs inhomogeneous simulation profiles to compute either $\rho_\mathrm{eq}$ or $c_1$, which depend
on the imposed external potential.%

\subparagraph{Training}

The necessary training data is collected by performing grand-canonical Monte Carlo simulations of a Lennard-Jones system
above the critical temperature at different chemical potentials. From these configurations the radial
distribution function $g(r)$ and $h(r) = g(r)-1$ are extracted. For the simulation bulk DCF $c_2(r)$, one needs to solve 
the bulk Ornstein--Zernike equation
\begin{equation}
c_2(r) = \frac{1}{2 \pi^2} \int \frac{\sin(kr)}{kr} \left( \frac{\hat{h}(k)}{1+\rho_b \hat{h}(k)} \right) k^2 \odif{k} 
\end{equation}
where the hat symbol denotes Fourier transformed quantities.
Finally, the conversion from radial to planar geometry needs to be performed on $c_2(r)$ as described earlier, see eq.~(\ref{eq:omega_rtoz}).

The (scalar) free energy functional is approximated by a convolutional neural network with periodic and dilated convolutions, where the latter
means increasing the reach of the kernel by stretching and padding with zeros, without adding more parameters to the model. The input density profiles are discretized on $320$ points,
the convolution kernel size is $3$, the dilation factor $2$ and the number
of layers $6$, with $[16,16,32,32,64,64]$ channels for the individual layers.

Applying auto-differentiation to the scalar output twice produces
a Hessian $\pdv{F^{(2)}_\theta}/{\rho_i,\rho_j}$ of rank two. In order to limit the computational cost, a randomly sampled subset of $10$ rows was used
for every loss evaluation, defined by
\begin{equation}
L(\theta) = \sum_{i,j} \left( \int_{|z_i-z_j|}^\infty \odif{r} 2 \pi r c_2(r) - \bar c_2(\theta, |z_i - z_j|)   \right)^2
\end{equation}

By fixing the second derivative of the free energy to a certain value one is still left with two ``integration constants'' in the parametrization of $F_\mathrm{ex}$. The first
one is determined in a consistent way by adding a regularization loss that constrains the first derivative of the free energy to be equal to
the excess chemical potential $\mu_\text{ex}$ (which is
the direct correlation function $c_1$ (times $k_B T$) for bulk densities), i.e.\
\begin{equation}
	L_\mathrm{reg}(\theta) = \frac{1}{n}\sum_i \left( \frac{1}{A \Delta z} \pdv{F^{(2)}_\theta}{\rho_i} - \mu_\text{ex}(\rho_b) \right)^2
\end{equation}
where $\beta \mu_\text{ex}(\rho_b) = \beta \mu - \ln (\lambda^3 \rho_b)$ and $\rho_b$ is the measured bulk density in the grand canonical simulations
at chemical potential $\mu$.
The second integration constant can be determined by the fact that $F_\mathrm{ex}[0] = 0$, i.e.\ by subtracting the value $F^{(2)}_\theta[0] = C$ from the
neural functional.

After the training is completed the resulting functional $F^{(2)}_\theta$ can be used to compute equilibrium profiles
for arbitrary (planar) external and chemical potentials.
Similar to Ref.~\cite{sammuller2023neural}, the network was never trained
on self--consistent density profiles. Therefore it is  a priori unclear whether self--consistent, converged inhomogeneous equilibrium profiles can be determined and are
accurate, especially since during training the network encountered only bulk systems.

\subparagraph{Results}

The method is compared to two other functionals and Monte Carlo simulations, which serve as ground truth. The first functional $F_\mathrm{MF}$, is an analytical
mean-field approximation of the Lennard-Jones potential using fundamental measure theory for the hard-sphere reference system. The other is an identically
parametrized neural network, trained on the output of the \emph{first} functional derivative of $F_\mathrm{ex}(\theta)$,
essentially a version of the method described in Ref.~\cite{sammuller2023neural}, leading to a free energy $F^{(1)}_\theta$.  

A variety of quantities can be compared in order to assess the quality of the learned functional approximation. Of special interest are the resulting equilibrium density profiles for varying external and chemical potentials,
the equation of state and the value of the free energy itself. The general observation is that $F^{(2)}_\theta$ performes comparatively or better
than $F^{(1)}_\theta$, especially (far) outside the training region, where the other methods
often fail to converge.
Ensuring numerical stability and self-consistent fixed points outside the training region is notoriously difficult for parameterized free energy functionals, probably due
to unwanted overfitting to the training data. Conditioning on a higher--order correlation function seems to be a beneficial strategy in this regard.

In summary, pair-correlation ($c_2$) matching is able to learn a very stable and accurate
neural density functional without having ``seen'' during training inhomogeneous density profiles due to an explicit external potential %
(other than the inter-particle potential through the radial
distribution function). Taking advantage of the convertability between
radial and planar geometry the network could be trained on the computationally simpler to handle geometry without losing information about the radial components.
Nevertheless, this approach rests on the fact that DCFs are available through the solution of the Ornstein-Zernike relation, which especially in the case of anisotropic
pair potentials becomes a highly non--trivial task.

\subsection{Learning the map $\rho \leftrightarrow \vext$}
\label{sec:learning-map}

\subsubsection{Bayesian methods}
The methods described in Refs.~\cite{yatsyshin2021, malpica2023physics} are based on Bayes' theorem, namely
\begin{equation}
    \label{eq:bayes_thm}
    P(Q| \Dcal) \propto P(Q) P(\Dcal | Q) \,,
\end{equation}
where $\Dcal$ is short for the observed data, $Q$ are the model parameters and $P(Q|\Dcal)$ is the posterior 
distribution of $Q$ given $\Dcal$. 
While both works are based on the same statistical method, they differ in their aims. The approach in Ref.~\cite{yatsyshin2021} focuses on learning the free energy functional of a 1D hard rod fluid, while in Ref.~\cite{malpica2023physics} the mapping of $\rho(x) \to V^\mathrm{ext}(x)$ is investigated.
Here we limit ourselves to the later work as the methods used in both are rather similar.

The principal question formulates a typical inverse problem:
\emph{Given a density profile $\rho(\rv)$ (i.e.\ the data $\Dcal$ in the Bayesian sense) resulting from an unknown external potential $V^\rmext$ is it possible to
infer the potential (i.e.\ its parameters $Q$ along with their uncertainty) given the density profile alone?}
The answer to this is yes and it is given by the distribution over $Q$, i.e.\ $P(Q| \Dcal)$. The interesting point is that 
$P(Q| \Dcal)$ contains the uncertainty in the final solution for the external potential; this is relevant if the input data would come from noisy simulations, say. 

The authors start by parameterizing the external potentials using a set of parameters $\{Q_i, Q'_i\}$, with $i \in [1, \ldots, i_\mathrm{max}]$
\begin{equation}
    \label{eq:Vext_q}
    V(z) = \sum_{i=1}^{i_\mathrm{max}} Q_i \exp \left[ - (z-z_i)^2 / \exp Q'_i \right]
\end{equation} 
where $z_i$ are points in the simulation box, corresponding to the center of each Gaussian. The choice of parameterization is essentially arbitrary but a 
good compromise between simplicity and expressiveness is helpful for training and generalization.
In order to determine $P(Q|\Dcal)$, the two distributions on the right hand side of Eq.~(\ref{eq:bayes_thm}) need to be specified.
The prior distribution $P(Q)$ is somewhat arbitrary and chosen to be Gaussian with zero mean and a diagonal covariance matrix, 
    $P(Q) = \Ncal(0, \Sigma_Q)$. 
It remains to specify the quantity $P(\Dcal|Q)$, which tells us how probable a certain density distribution is
given the parameters $Q$ of the external potential.
For this, the probabilistic interpretation of the particle density itself is used:
\begin{equation}
    \label{eq:likelihood}
    P(\Dcal|Q) = \prod_i \rho(z_i | Q)\;,
\end{equation}
where one multiplies the probabilities over all space points.
This mapping of the parameters $Q$ to the density profile is not analytically known, but numerically through simulations (in general) or through cDFT using the exact functional in the specific case of hard rods considered here.
In Markov chain Monte Carlo simulations, the $Q$ are included in the sampling and a detailed balance criterion is formulated 
which generates a histogram according to the  distribution of \cref{eq:bayes_thm}. %
From this histogram the final $\vext$ follows from the most probable values of $Q$, and a band of uncertainty in $\vext$ can be drawn from the quantiles.

\begin{figure}
    \centering
    \includegraphics[width=.9\linewidth]{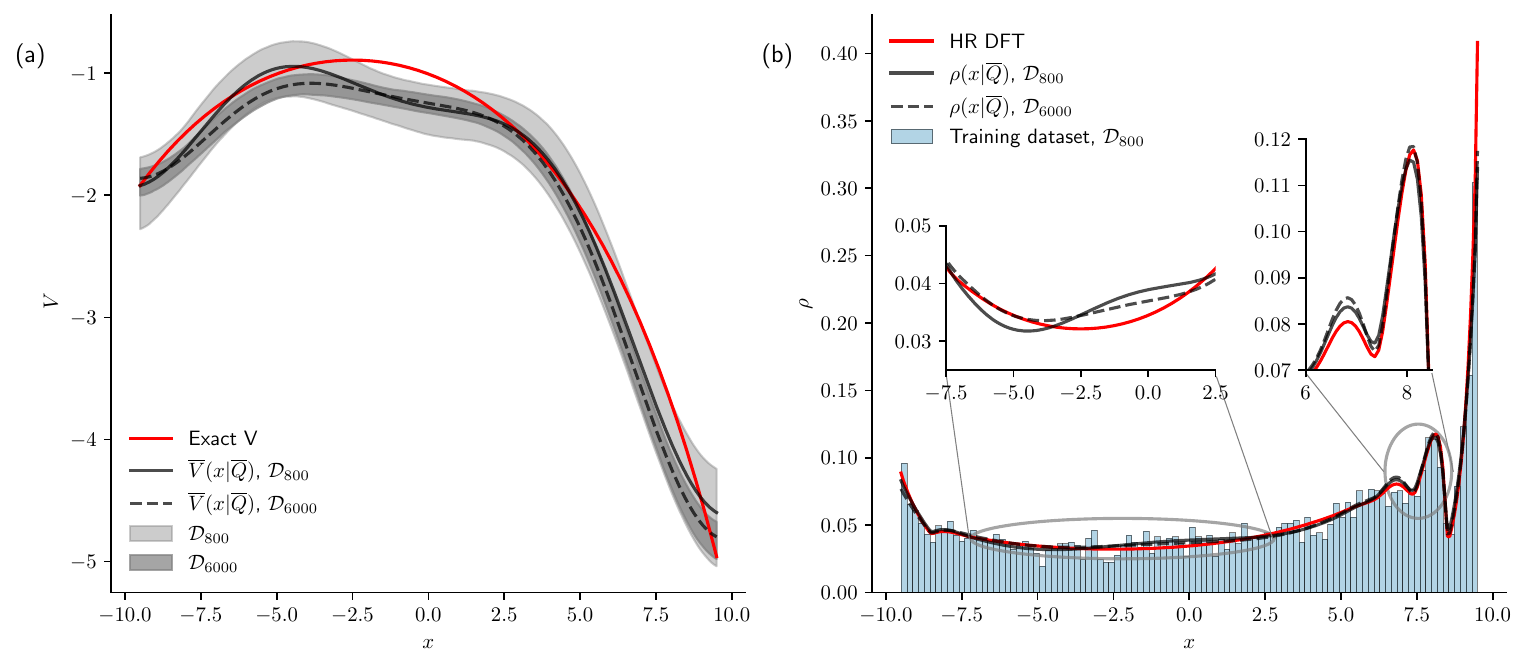}
    \caption{Predicted external potentials (a) and corresponding density profiles (b)
    for two different dataset sizes of \num{800} and \num{6000} with 99\% confidence intervals. The differences
    in the density profiles are smaller than those in the potential representation. Here the 1D coordinate is denoted by $x$. Figure taken from Ref.~\cite{malpica2023physics}.}
    \label{fig:bayes}
\end{figure}

\subparagraph{Performance}
In order to test the approach it was trained with sets of three different sizes (\numlist{800;2400;6000}) with configurations 
coming from Monte Carlo simulations.
With increasing dataset size the histograms for the individual parameters $Q_i$
become narrower, signaling an increased confidence in the parameter.
This also carries through to the prediction of the external potential, whose confidence intervals gets smaller with increased
training set size.
As for the maximum a posteriori (MAP) prediction of the potentials, one observes slight discrepancies between the ground truth and the prediction of the external
potentials. When looking at the corresponding density profiles the difference 
between ground truth and predicted value becomes smaller, signaling a certain
freedom in the choice of external potentials, see Fig.~\ref{fig:bayes}. (This appears to be similar to the problem of finding the exact excess free energy functional for hard rods in Ref.~\cite{shang2020FEQL}.) 
Nevertheless the exact $\vext$ is for the largest part contained in the 99\% probability region, even for the smaller training set size of \num{800}.
Indeed it was observed that the minimal training dataset size at which
the predictions fail to recover the results of the small training set is around
\num{360}.

\subsubsection{Gaussian Process}
\label{sec:guassian_proc}

Ref.~\cite{fang2022reliable} addresses the map from the external potential to the density profile. It will be represented by a typical surrogate model of machine learning, namely Gaussian processes (GP), which replaces the cDFT calculation or the direct simulation. The authors exemplify the method for the 1D hard rod system, for the sake of laying out the principle. A particular focus lies in formulating a reinforcement learning scheme, i.e. in the automated inclusion of new training data if the output of the surrogate model for a particular new external potential is of potential low accuracy. 

The input for the map $\vext \to \rho$ is discretized with $p$ points on the Cartesian $z$ axis, forming the vector
\begin{equation}
    \xv = \mu - V^\rmext(z) = \psi(z) \in \reals^p\;,
\end{equation}
and the resulting output is the discretized density vector $\rhov \in \reals^p$ with $\rho_i = \rho(z_i)$.
For multiple inputs ${\xv^{(1)},\ldots, \xv^{(n)}}$ we have the density outputs ${\rhov^{(1)},\ldots, \rhov^{(n)}}$, and the assumption of the GP model is that the vector of densities $\hat\rhov_j$ \textit{at one space point $j$} follows a multivariate normal distribution 
\begin{equation}
    \hat \rhov_j = [\rho_j^{(1)}, \rho_j^{(2)}, \ldots, \rho_j^{(n)}]^\top \sim \mathcal{MN}(\muv_j,\sigma^2_j \Rv^{(n)})
  \label{eq:gp1}
\end{equation}
where $\Rv^{(n)}$ is the $n \times n$ correlation matrix between densities at a fixed
grid point with elements $ R^{(n)}_{ij} = K(||\xv^{(i)}, \xv^{(j)}||)$. It is through this kernel function that the density at one grid point  depends on the external potential at different grid points. In Ref.~\cite{fang2022reliable}, so-called Mat\'ern kernel functions were used, which depend on a range parameter $\gamma$ that needs to be estimated.
The point is now to consider the $n$ densities $\rhov^{(1)}(\xv^{(1)}),...,\rhov^{(n)}(\xv^{(n)})$ as input data $\Dcal$ from simulations, say. One adds a new external potential $\xv^{(n+1)}=\xv^*$ with unknown density output $\rho_j^{(n+1)}=\rho_j^*$ and is interested in the distribution $P(\rho_j^*|\Dcal)$ given the input data. Since the joint distribution
$P(\rho_j^{(1)},...,\rho_j^{(n+1)})$ still follows the multivariate distribution (\ref{eq:gp1}), one has
$P(\rho_j^*|\rho_j^{(1)},...,\rho_j^{(n)})=P(\rho_j^{(1)},...,\rho_j^{(n+1)})/P(\rho_j^{(1)},...,\rho_j^{(n)})$ which is still Gaussian but depends on the means $\muv_j$ and variance parameters $\sigma^2_j$. Marginalizing out these with certain assumptions on their prior distributions finally leads to a closed form for $P(\rho_j^*|\Dcal)$.

The remaining challenge is to find a suitable strategy for
choosing an optimal set of inputs for $\hat \rhov_j$ and estimating the range parameter $\gamma$ in the kernel function $K$.
The proposed architecture is termed ALEC (active learning with error control) and functions as an adaptive emulator in the following way.
After supplying the first batch of training profiles the internal state of the estimator is built,
i.e.\ the correlation matrix and the corresponding guesses for $\theta$ and $\sigma$ are computed. When a new test input $\xv^*$ (external potential) is given to the system, it predicts the output
$\rho_j(\xv^*)$ by sampling from the appropriate probability distribution with mean $\hat{\theta}$ and variance $\hat{\sigma}$, both depending on the
covariance between samples in the training set. The parameter $\gamma$ is estimated such that the performance is maximized.
Since the predictive distribution $P(\rho^*_j|\Dcal)$ is known analytically, the predictive error can be directly assessed.
In case the variance exceeds a certain threshold, the systems generates new training data for $\xv^*$ 
and includes it into its training set (augmented set).
Since the training set is now increased by one, it is necessary to recompute the covariance matrix $\Rv^{(n+1)}$ and its inverse. Ref.~\cite{fang2022reliable} discusses methods to speed up this process. For benchmarking, ALEC is compared to training setups with a random choice of training samples and a different adaptive scheme (D-optimality) and shown to perform better in all three cases.

Although interesting in terms of the adaptive learning strategy, the approach is perhaps the most remote of the discussed methods to address fundamental or applied problems in cDFT. The map $\vext \rightarrow \rho$ is still the easiest, also for simulations to generate ground truth data. A surrogate model for simulations would be needed in perspective only for very costly simulations, such as  complex biomolecules with many internal degrees of freedom.

\section{Outlook to related problems}
\label{sec:quantum_pft}

\subsection{Liquid state theory}
\label{sec:ie}

A recent review article \cite{wu2023perfecting} gives an overview on recent advances in liquid state theory using ML methods  which we recommend to the reader and therefore restrict ourselves to a few points only.

The integral equation (IE) approach to liquid state theory aims at determining the direct correlation function $c_2(x,x')$ and the total correlation function $h(x,x')$ which are linked by the Ornstein--Zernike (OZ) equation (\ref{eq:oz}). A second equation is needed which is usually termed \textit{closure equation} and which for a fluid with intermolecular pair potential $\phi(x,x')$ takes the form \cite{hansen2013theory}
\begin{equation}
  \label{eq:closure}
  \ln( h(x,x') +1) + \beta \phi(x,x') = h(x,x') -c_2(x,x') + b(x,x')\;.
\end{equation}
The unknown function $b(x,x')$ is called the bridge function. In terms of the Mayer expansion it has a clear diagrammatic definition which, however, does not allow for analytic resummations, even approximate ones. For simple liquids with repulsive cores it has been noted that the approximation of $b$ by a $b^\text{ref}$ from a reference system system (usually hard spheres) gives good results \cite{rosenfeld1979universality} (bridge function universality). Empirical closures have expressed $b$ pointwise as functions of $h$, $c_2$ and $\beta \phi$ \cite{hansen2013theory}. Recent work \cite{lee2021bridge} has investigated an ML closure for simple fluids where $b$ is a pointwise function of $h$, $c_2$, the derivative of $h-c_2$ and the fluctuations of the pair correlation function $g=h+1$. Indeed it has been found that a variant of bridge function universality holds approximately, i.e. that for ``hard'' potentials (steeply repulsive near the origin) one approximate ML closure is found while for ``soft'' potentials (weakly diverging or permitting core overlap) another one is found. 

An explicit connection to cDFT arises if one considers the cDFT derivation of the closure Eq.~(\ref{eq:closure}) \cite{rosenfeld1993,oettel2005,borgis2021accurate}. If the excess free energy functional is functionally expanded around a (homogeneous or [if $\vext \neq 0$] inhomogeneous) reference density profile and the grand potential is minimized in the additional presence of a test particle ($\vext \to \vext+\phi$), then Eq.~(\ref{eq:closure}) arises and $b(x,x')$ is the functional derivative of the Taylor expanded excess functional from which all terms up to second order are subtracted. Thus ML representations of the excess free energy functional would also contain an approximate solution to the IE problem.

\subsection{Electron DFT}

A dedicated and more comprehensive review on the connection between ML and electron DFT is in the chapter on  \textit{Machine learning in quantum density functional theory} by Thorsten Deilmann in this book \cite{deilmann2024}. Here we limit ourselves to discussing ML assisted functional building approaches in electronic DFT and relate these to the cDFT approaches, if possible. 
In contrast to cDFT, in quantum DFT (qDFT) only one system is of major interest: interacting electrons with their pairwise Coulomb repulsion in the presence of nuclei. The Hohenberg--Kohn proof \cite{hohenberg1964inhomogeneous} entails that there exists a unique \textit{energy} functional $E[n(\rv)]$ where the electron density is denoted by $n(\rv)$. Conventionally (in the Kohn-Sham approach) it is split 
\begin{equation}
   E[n] = K_S[n] + E^\text{ext}[n] + E_H[n] + E_\text{XC}[n]
\end{equation}
where 
$K_S[n]=-\hbar^2/2m  \sum_i \int d\rv \phi^*_i[n]  \nabla^2  \phi_i [n] $
is the kinetic energy of noninteracting electrons in occupied Kohn--Sham orbitals $\phi_i$ \cite{sham1965}. 
$E^\text{ext}[n]= \int d\rv n(\rv)\vext(\rv)$ is the external energy due to the external potential $\vext$ exerted by the nuclei. 
$E_H[n]=(e^2/8\pi\epsilon_0) \int d\rv \int d\rv' n(\rv) n(\rv')/|\rv-\rv'|$ is the Hartree energy which is the classical interaction energy 
of an inhomogeneous electron distribution and which is of typical mean-field form. 
The quantum--mechanical exchange and correlation effects are buried in the \textit{exchange--correlation} functional $E_\text{XC}[n]$ 
whose exact form is unknown; it is the equivalent of the excess free energy functional $\FE^\text{ex}[\rho]$ minus the mean--field part in cDFT. 
Once a specific form for $E_\text{XC}[n]$ is assumed, a typical iteration scheme for finding the minimum of $E$ proceeds via the solution of a one--electron Schr\"odinger equation in an external potential (Kohn--Sham potential $v_S[n_j]$) which depends on the electron density $n_j$ in the previous step $j$, 
\begin{eqnarray}
  \left(  - \frac{\hbar^2}{2m} \nabla^2 + v_S[n_j] \right) \phi_i(\rv) &=& \epsilon_i \phi_i(\rv) \label{eq:ks1} \\
  n_{j+1}(\rv) &=& \sum_i | \phi_i(\rv)|^2 \quad \text{\small{(occupied orbitals)}} \label{eq:ks2}
\end{eqnarray}
and the construction of the density  $n_{j+1}$ in the next step $j+1$ through the sum of the densities in occupied orbitals (fixed by the total 
number $N$ of electrons). Hereby the Kohn--Sham potential is given by $v_S[n] = \vext + \delta(E_H+E_\text{XC})/\delta n$
(thus $-\delta(E_H+E_\text{XC})/\delta n$ is the equivalent of $c_1$ in cDFT).
The procedure is repeated until the density is converged. This iteration is the equivalent of the iterative solution (\ref{eq:it1},\ref{eq:it2}) of the fixed point Euler--Lagrange equation (\ref{eq:el}) in cDFT. 

ML methods in qDFT can be used in similar ways as in cDFT. In difference to classical systems, ground truth data do not come exclusively from dedicated simulations (like quantum Monte--Carlo \cite{ceperleyAlder1982qMC,anderson2007}) owing to numerical challenges. For test problems in 1D and few electrons, the many--body Schr\"odinger equation can be solved exactly. For ``real world'' problems, data sets from experiment or advanced (but laborious) and more systematic electron structure calculations can be used (see e.g. the Main
Group Chemistry DataBase (MGCDB84) \cite{30yrsDFT2017}) which are inherently approximative but deemed more precise than the average qDFT result.  

Calculations in 1D (mostly for electrons with an exponentially screened potential) have been a test bed for ML methods. One can introduce a local energy per particle $e(\rv)$ through $E= \int d\rv n(\rv) e(\rv)$. Nonlocal mappings
$n \leftrightarrow e[n]$ can be represented by standard convolutional networks or MLPs. Examples for this can be found in Refs.~\cite{nagai2018,schmidt2019,lili2021} which differ in the forms of their loss function (taking into account the energy density, its spatial derivative, and the density distribution). An obvious choice is to use the density $\rho^\star$ and energy density $e^\star$ from converged Kohn--Sham (KS) iterations (with an ML functional) in the loss function to guarantee that the ML functional delivers self--consistent solutions (this is similar to Refs.~\cite{cats2021,simon2024machine} in cDFT). However, large training data sets are needed and together with the numerical costs of the KS iterations some doubts to the scalability to interesting 3D systems arise. An interesting proposal for a partial remedy is in Ref.~\cite{lili2021} which suggests to use also intermediate energies and densities from the KS iterations in the loss function and training process to multiply the training data.    

Nevertheless, ML functionals for realistic 3D systems (i.e.\ molecules) have been published recently by Google collaborations \cite{kirkpatrick2021,lili2022}. The Deepmind21 (DM21) functional of Ref.~\cite{kirkpatrick2021} addresses in particular the problem of correct dissociation curves of molecules by approximately solving the fractional charge and spin problem of DFT. The price to be paid is the use of approximate density distributions in the training and training/evaluation of the ML functional in a non--selfconsistent, perturbative  manner. As for the other work, the idea followed in Ref.~\cite{lili2022} is to learn a symbolic (analytic) exchange--correlation functional based on a reduced set of analytical terms and operations which is motivated by the structure of the most reliable ``human--made'' exchange--correlation functionals. The analytic form allows for self--consistency in all functional evaluations. The authors claim a superior description of their GoogleAcceleratedScience22 (GAS22) functional across the MGCDB84 database. In methodology, there is a strong similarity to the Functional Equation Learner of Ref.~\cite{shang2020FEQL} for cDFT (see Sec.~\ref{sec:eql}).

The Kohn--Sham approach to qDFT via KS orbitals and the self--consistent iteration of Eqs.~(\ref{eq:ks1},\ref{eq:ks2}) is a quite special trick to approximate the kinetic energy $K[n]$ in qDFT. In cDFT, the kinetic energy of particles is buried in the ideal part $\FE_\text{id}[\rho]$ of the free energy, and the corresponding free energy density is a simple local functional. This is not so in qDFT, but it would seem natural to search for an explicit kinetic energy functional $K[n]$ here as well (``orbital free DFT''). In fact this was the subject of early work on finding functionals for 1D electrons with ML \cite{burke2012,burke2016} where the kinetic energy functional $T[n]$ was represented using a rather simple kernel ridge regression. The obtained functionals still suffered from a comparatively poor representation of the functional derivative $\delta T/\delta n$. These problems had been solved including a proper training on the functional derivative \cite{hauser2020} and substituting the numerically expensive kernel ridge regression by more effective convolutional networks representing the map to the energy per particle $n \to e[n]$. It remains to be seen whether these insights for 1D will translate to a workable 3D orbital--free ML functional.     

\subsection{Power functional theory}

We have attempted to show the similarity of the ML problem in cDFT and qDFT (despite the differences in the specific treatments) which is simply a consequence of the existence of the functional map $\rho \leftrightarrow \FE[\rho]$ or $n \leftrightarrow E[n]$. If the general (classical or quantum--mechanical) nonequilibrium problem could be formulated with a similar functional map, one would expect that ML techniques can be of similar use. For classical systems, a functional formulation, power functional theory (PFT), has been proposed about 10 years ago which shows a close resemblance to cDFT and suggests itself for extending the ML techniques. The current status of PFT is reviewed in Ref.~\cite{schmidt2022review}. Here we restrict ourselves to the case of Brownian (overdamped) dynamics which is of superior relevance in the colloidal domain and describes the thermal motion of macroobjects in a solvent. Specifically we consider isotropic particles for which the force on a single particle $i$ is proportional to its velocity, ${\mathbf f}_i=\gamma \vv_i$. 

The central quantity in Brownian PFT is the ensemble-averaged one--particle current $\Jv(\rv,t) = \langle \sum_i \delta(\rv -\rv_i) \vv_i \rangle$ (where 
the average is over initial conditions with a prescribed $n$--particle distribution function in space). The space and time dependent density distribution is linked to the current via the continuity equation $\dot \rho(\rv,t)=-\nabla \cdot \Jv(\rv,t)$. It can be shown that there exists a functional $R[\Jv]$ 
such that the physical nonequilibrium single particle current $\Jv_\text{neq}(\rv,t)$ is determined by functional minimization
\begin{equation}
  \label{eq:minimizationPFTbd}
        \left.\frac{\delta R[\Jv]}{\delta {\Jv(\rv,t)}}\right\vert_{\Jv=\Jn} = 0 \quad .
\end{equation}
For the original proof see Ref.~\cite{schmidt2013power}, some mathematical issues have been clarified and corrected in Ref.~\cite{lutsko2021}.
Similar to the grand potential functional in DFT, $R[\Jv]$ is split into an intrinsic part and a part accounting for the interactions with a space and time dependent external potential $\vext(\rv,t)$:
\begin{equation}
R[\Jv]  =: R^\text{int}[\Jv] + \int  \dd \rv \; \Jv \cdot \nabla \vext
    + \int  \dd \rv \; \rho {\dot V}^\text{ext} \;.
\end{equation}
  From Eq.~(\ref{eq:minimizationPFTbd}) then follows an Euler-Lagrange equation of the form 
  $\frac{\delta R^\text{int}}{\delta{\Jv}} = - \nabla \vext \;.$
This internal part $R^\text{int}[\Jv]$ (like the free energy functional $\FE[\rho]$ in cDFT) is a unique functional \textit{independent of the external potential} and can be split further by the $\textit{ansatz}$:
\begin{equation}
 R^\text{int}[\Jv] = \int  \dd \rv\; \Jv \cdot \nabla \frac{\delta \FE[\rho]}{\delta \rho} + \Pid[\Jv] + \Pex[\Jv] \; .
\end{equation}
$\Pid= \gamma  \int  \dd \rv \; \Jv^2/(2\rho)$ is the ideal gas part of the functional for the dissipated power, and
$\Pex$ is the (generally unknown) excess part. Note that $R^\text{int}$ is strictly a functional of the current $\Jv$ only, and densities $\rho(\rv,t)$ are determined through the continuity equation and only depend on the current $\Jv(t'<t)$ at earlier times.
With these definitions, Eq.~(\ref{eq:minimizationPFTbd}) becomes an implicit equation for
the exact non-equilibrium current $\Jn$:
\begin{equation}
    \gamma  \Jn  = - \kb T  \nabla\rho -\rho \nabla \vext \underbrace{
                    -  \rho \nabla \frac{\delta \FE_\text{ex}[\rho]}{\delta \rho} -   \rho \left.  \frac{\delta \Pex}{\delta \Jv} \right|_{\Jv=\Jn}}_{-\rho \delta R^\text{int}_\text{ex}/\delta \Jv} \;.
\end{equation}
The first two terms here give the exact nonequilibrium current of an ideal gas in an external potential. With the third term added, dynamic DFT \cite{tevrugt2020ddftreview} is recovered in which the dynamics solely depends on the \textit{instantaneous} density profile $\rho(\rv,t)$ through the equilibrium free energy functional. The fourth term with the functional derivative of the excess power corrects the quasi--equilibrium approximation of dynamic DFT to recover the full nonequilibrium current. The sum of the third and fourth term can be viewed as the functional derivative of the excess part of the internal part of the functional $R^\text{int}$. 

In view of this, the full classical equilibrium and (Brownian dynamics) nonequilibrium many--particle problem is the problem of finding the two unknown excess functionals $\FE_\text{ex}[\rho]$ and $\Pex[\Jv]$. As described earlier, over the past decades knowledge about $\FE_\text{ex}[\rho]$ has been accumulated and has entered the diverse ML approaches described in Sec.~\ref{sec:approaches}. This is different in the case of $\Pex[\Jv]$: currently we do not know an exact excess power functional for a simple (even 1D!) model system, and knowledge on the general structure of $\Pex[\Jv]$ has only begun to be gathered \cite{schmidt2022review}.           

The problem of a nonequilibrium steady state ($\Jv=\text{const.}$) is an excellent example for which the relevance of $\Pex[\Jv]$ has been demonstrated and for which the aforementioned full problem has been tackled with ML methods \cite{delasheras2023}. The system is 3D Lennard--Jones in a box with periodic boundary conditions to which an external force has been applied which produces a steady flow in the $z$ direction with constant current $J_z$. The ensemble--averaged density $\rho(z)$ and velocity $v_z(z)$ are inhomogeneous but linked through $J_z=\rho(z) v_z(z)$. With dedicated Brownian dynamics simulations, the averaged internal force density
\begin{equation}
 \mathbf{F}_\text{int}(\rv,t) = -\langle {\textstyle \sum_i} \delta(\rv-\rv_i) \nabla_i u(\rv^N) \rangle = - \rho \frac{\delta R^\text{int}_\text{ex}}{\delta \Jv}
\end{equation}
can be obtained as ground truth. This internal force density is the density profile times the negative of the functional derivative of the excess part of $R^\text{int}[J]$ (it contains both contributions by $\FE_\text{ex}[\rho]$ and $\Pex[\Jv]$) and thus offers itself for an ML representation. For the steady state problem, a functional dependence on the time--independent $\rho$ and $\Jv$ is needed but since simulations only realize constant currents it is better to include the velocity $\vv$ as a functional variable.    The authors of Ref.~\cite{delasheras2023} proceeded to represent the force per particle $\mathbf{f}_\text{int}=\mathbf{F}_\text{int}/\rho$ as a convolutional network $f_{z,\text{int}}^\star(z)[\rho(z'),v_z(z')]$ where the density and velocity distribution around a point $z$ were mapped to the force at $z$. From the results, it is gratifying to see that the ML functional representation of 
$\mathbf{f}_\text{int} = - \delta R^\text{int}_\text{ex}/\delta \Jv$ indeed worked \textit{independent of the external force}. Furthermore, principal shortcomings of the dynamic DFT approximation could be highlighted. Conceptually and methodwise, there is a strong similarity to Refs.~\cite{sammuller2023neural,sammuller2023neural_ii} (from the same group) which tackled the ML representation $c_1^\star[\rho]$ of the excess free energy functional derivative $ - \beta \delta \FE_\text{ex}/\delta\rho$ in equilibrium cDFT, see the discussion in Sec.~\ref{sec:nn-mapping}.

\section{Summary and conclusion}

We have discussed in some detail the recent developments in classical density functional theory connected with the use of machine learning techniques. Here, we laid particular focus on work which addressed the maps between density distribution and free energy functional (or its first derivative) and density distribution and external potential. We have not treated the use of machine learning techniques to classify and interpret results of high through--put DFT calculations \cite{wu2022} which is a topic which appears to be particularly relevant in quantum DFT and materials science (see also the chapter on  \textit{Machine learning in quantum density functional theory} by Thorsten Deilmann in this book \cite{deilmann2024}). 
Most of the tested techniques (explicit parametrizations of functionals, multi-layer perceptrons, analytical learning schemes, Bayes techniques) have been applied to the one-dimensional hard rod system where the exact functional is known. One  can fairly say that these demonstrations were successful and the representability of functional maps using ML is possible with quantitative accuracy. However, to date there have been limited applications to more realistic, three--dimensional systems with truly novel results. This will certainly constitute a challenge for the future. Possible roads for the future are (i) the standard inclusion of a $\rho \rightarrow c_1[\rho]$ training in simulations as an efficient way to organize the simulation database \cite{sammuller2023neural}, (ii) further elucidations of the analytic structure of functionals in 3D systems \cite{cats2021,simon2024machine} and (iii)  accounting for uncertainties in machine--learned representations \cite{malpica2023physics}. It will be interesting to see whether machine--learning supported cDFT  can expand to systems where currently there are little analytical insights into functionals, such as particles with internal degrees freedom (which are not decoupled), particles of complex anisotropic shape or biomolecules.  
Regarding functional building and the use of machine learning, we have noticed a strong conceptual overlap to quantum DFT, in particular with regard to orbital--free approaches. The problem setting in classical DFT is however much more diverse: While in quantum DFT only the system of interacting electrons is of paramount importance, there are many Hamiltonians in the classical realm which deserve scrutiny and which need differing attention. In our outlook we also noted the great potential which lies in the functional formulation of classical many--body dynamics which then can be tackled similarly to equilibrium DFT problems using methods of machine learning.

\begin{acknowledgement}
We gratefully acknowledge funding by the Deutsche
Forschungsgemeinschaft (DFG, German Research Foundation) under Germany’s Excellence Strategy EXC no. 2064/
1, Project no. 390727645.
\end{acknowledgement}

\printbibliography
\end{document}